\RequirePackage{fix-cm}
\documentclass[smallextended]{svjour3b}       
\smartqed  
\usepackage{graphicx}
\def\bibentry#1{\hangindent=2em #1}

\def\beq{\begin{equation}}
\def\eeq{\end{equation}}
\def\bea{\begin{eqnarray}}
\def\eea{\end{eqnarray}}

\newcommand{\gev}{\, {\rm GeV}}
\newcommand{\tev}{\, {\rm TeV}}

\def\mweak{m_{\rm weak}}

\begin{document}

\title{Higgs naturalness and the scalar boson proliferation instability problem}


\author{James D. Wells \\
             {\it Physics Department \\
             University of Michigan\\
             Ann Arbor, MI 48104 USA}
}



\institute{Published in {\it Synthese}. DOI: 10.1007/s11229-014-0618-8}

\date{Presentation at the Interdisciplinary Workshop on the Epistemology of the LHC, University of South Caroline (25-26 April 2014).}

\authorrunning{Wells}
\titlerunning{Higgs naturalness}

\maketitle

\begin{abstract}
Sensitivity to the square of the cutoff scale of quantum corrections of the Higgs boson mass self-energy has led many authors to conclude that the Higgs theory suffers from a naturalness or fine-tuning problem. However, speculative new physics ideas to solve this problem have not manifested themselves yet at high-energy colliders, such as the Large Hadron Collider at CERN. For this reason, the role of naturalness as a guide to theory model-building is being severely questioned. Most attacks suggest that one should not resort to arguments involving gravity, which is a much less understood quantum field theory. Another line of attack is against the assumption that there exists a multitude of additional heavy states specifically charged under the Standard Model gauge symmetries. Nevertheless, if we give ground on both of these assaults on naturalness, what remains is a naturalness concern over the prospect of numerous additional spin-zero scalar states in nature. The proliferation of heavy scalars generically destabilizes the Higgs boson mass, raising it to the highest and most remote scalar mass values in nature, thus straining the legitimacy of the Standard Model. The copious use of extra scalars in theory model building, from explaining flavor physics to providing an inflationary potential and more, and the generic expectation of extra scalar bosons in nature argues for the proliferation instability problem being the central concern for naturalness of the Standard Model.  Some approaches to solving this problem are presented.

\keywords{Higgs Boson \and  Naturalness \and Fine-tuning \and Hierarchy Problem}
\end{abstract}

\vfill\eject
\tableofcontents


\section{Introduction}
\label{intro}

The discovery of the Higgs boson (Aad 2012, Chatrchyan 2012) and nothing else exotic so far (Soni 2013) has put to rest questions of the existence of the Higgs boson, and rejuvenated questions about its viability without additional dynamics beyond the Standard Model. The Higgs boson is unique among the elementary particles in that its quantum corrections are quadratically sensitive to high scales\footnote{For an historical overview of how understanding developed over time of the quantum corrections of scalars and the Higgs boson see Giudice (2008).}. This leads to what many perceive to be a naturalness problem for the Higgs boson.  

To be more precise, if we compute in quantum field theory the self-energy of the Higgs boson field, we find that the Higgs mass is
\beq
\label{eq:bare}
m^2_H=m_{\rm bare}^2+\frac{y_t^2}{16\pi^2}\Lambda^2 + \delta{\cal O}(\mweak^2)
\eeq
where $m_H$ is the Higgs boson mass with measured value $125\gev$, $m_{\rm bare}$ is the Higgs boson bare mass parameter of the unrenormalized lagrangian, $y_t$ is the top quark Yukawa coupling with value close to 1, $\Lambda$ is the cutoff value of momentum in the top quark loop of the Higgs boson self energy
\footnote{Quantum loop computations of a scalar boson self-energy involve the integration of $\sim\int d^4q/q^2$ over internal loop momentum $q$ which is formally allowed to go to $\infty$. However the integral is quadratically divergent, meaning $\int d^4q/q^2\propto q_{max}^2\to \infty$. If we cut off the maximum value allowed for $q^2$ to be $q_{max}^2=\Lambda^2$, we say that $\Lambda^2$ is the ``cutoff value of the momentum" in the integral.}, 
and $\delta{\cal O}(\mweak^2)$ are all other quantum corrections, where $m_{\rm weak}$ is meant to designate the weak scale\footnote{When precision of speech is requested we can define $m_{\rm weak}=100\gev$. This scale is chosen parametrically to be close to the numerical values of the $W$ and $Z$ boson masses $m_W=80\gev$ and $m_Z=91\gev$, the top quark mass $m_t=175\gev$, the Higgs boson mass $m_H=125\gev$ and the Higgs boson vacuum expectation value $v=246\gev$, normalized to be the value around which the dynamical Higgs field is expanded in the lagrangian.}. Eq.~\ref{eq:bare} is explicitly highlighting the contributions to the Higgs boson mass from the top quark loop, but there are many more contributions.

The naturalness argument, 
first articulated by Susskind (1979), 
suggests that if the Standard Model is a valid theory up to a very high scale, say $\Lambda\sim M_{Pl}\sim 10^{18}\gev$, then $m_{\rm bare}^2$ has to be a very large and extraordinarily fine-tuned number to cancel\footnote{The bare mass-squared mass must be large and negative to cancel the ``infinite" part induced by the top quark. This is fine as long as the combination of the two terms is positive.}  the very large contribution $y_t^2\Lambda^2/16\pi^2$, thereby reproducing the small Higgs boson mass of $125\gev$. There is no equivalently disquieting equation in particle physics that apparently requires such dramatic fine-tuning of quantum corrections.  
Only the cosmological constant has perhaps more mystery of such large discrepancies compared to expectations
\footnote{The quantum field theory diagrams that contribute to the cosmological constant scale as the integral $\sim \int d^4q$ where $q$ is an internal particle momentum that is formally allowed to go to $\infty$. We can introduce a cutoff scale for this integral such that $\int d^4q=\Lambda_{CC}^4$, which we believe should be at least as large as $\sim M_Z^4$, since we believe we know the theory of nature at energy scales up to at least $M_Z$. This implies that the cosmological constant would naively be at least $\Lambda_{CC}^4> M_Z^4\simeq 10^8\,{\rm GeV}^4$, or perhaps even $M_{Pl}^4\sim 10^{72}\,{\rm GeV}^4$ if we allow our integral to be cutoff at the known scale of gravity $M_{Pl}=G_N^{-1/2}$, where $G_N$ is Newton's constant of gravity. This constitutes the quantum field theory expectation. However, measurements in cosmology tell us that $\Lambda_{CC}^4\simeq 10^{-47}\,{\rm GeV}^4$, in gross contradiction to our naive expectations. This is still an outstanding problem in physics. See Rugh \& Zinkernagel (2001) for a discussion.}. 
This problem sometimes also is called the ``hierarchy problem", in that there exists a large hierarchy of $10^{16}$ between the Planck mass and the weak scale, yet the quadratic divergences of the Higgs sector imply that the two scales should be similar. 

One can try to appeal to ``technical naturalness" (`t Hooft 1980), which states that a theory can have a technically Natural small parameter if a new symmetry emerges as the parameter goes to zero. For this reason the electron mass of $5\times 10^{-4}\gev$ is technically Natural despite being orders of magnitude smaller than the weak scale. This is because an electron chiral symmetry, where its left and right spins can transform independently, is recovered in the massless limit. There is no such recovered symmetry restoration when the Higgs mass goes to zero since the $H^\dagger H$ mass operator is invariant under all symmetries of the Standard Model and all chiral symmetries. Only a shift symmetry, 
whose transformation is defined to be $H\to H+{\rm constant}$, appears to be helpful, since $|H|^2$ and $|H|^4$ terms are disallowed by it. 
However, the shift symmetry is badly violated by the top quark Yukawa interactions $y_t QH t_R$, which is the origin of the offending $y_t^2\Lambda^2/16\pi^2$ term in Eq.~\ref{eq:bare}, and thus cannot protect the Higgs boson mass.

\section{Naturalness concerns}

The problem of naturalness as presented above is not without weakness. The core of the argument against naturalness being a serious problem is that there are no observables that cannot be accounted for in the theory. We always have infinities in quantum corrections that are formally cancelled by counter terms embedded in the bare parameter. Furthermore, if we regularize in dimensional regularization, artificially setting the number of dimensions to be $n=4-\epsilon$, the infinities show up as $1/\epsilon$ quantum corrections that are cancelled unceremoniously, in contrast to the seemingly dramatic cancellations of the $\Lambda^2$ corrections that arise in a cutoff regularization method. There is no culture or meaning of declaring that counter-term cancellations with $1/\epsilon$ corrections are outrageously fine-tuned. It is just a formal bookkeeping process to account for it, and all calculations can be matched to observables, and there is no conflict with the data. Naturalness, in this viewpoint, is unjustified hysteria generated by just one of our artificial means of keeping track of intermediate steps in a quantum field theory calculation. Other authors have addressed this viewpoint (Bardeen 1995, de Gouv\^ea et al.\ 2014, Farina et al.\ 2013). 

The above discussion has focused on quantum corrections in the pure Standard Model theory. The naturalness concern rears its head more confidently if we assume that there is new physics with unknown dynamics at a high scale $\Lambda$ that the Higgs boson couples to, which in turn generates quadratic sensitivities to $\Lambda$ in the quantum corrections of the Higgs boson mass. 

Sometimes it is argued that we know already there is a new scale, the scale of the onset of strong gravity and quantum gravity at $M_{Pl}$, and the Higgs boson mass is surely affected by dynamics there. However, it is not a solid argument that the Higgs boson mass must suffer from destabilizing quadratic corrections due to gravity alone. Indeed, there is no obvious separate shift symmetry violating interaction of the Higgs boson with gravity that is not already suppressed by powers of the $1/M_{Pl}$ coupling and the original Standard Model couplings. This only would leave corrections that are at most of order the Higgs mass. 
Furthermore, whatever non-perturbative concerns one might have for the Higgs boson inheriting instability up at the Planck scale due to gravity, it remains uncertain how to account for it. Quantum gravity is a notoriously unsolved mystery, and the naturalness issues of the cosmological problem being so small $10^{-47}\gev^4$ compared to expectations $10^{72}\gev^4$ further highlights our ignorance. It is plausible that any high-scale quantum gravity intuition that we might try to invoke is dramatically wrong, and so it is reasonable to question any quantum gravity argument  impugning Higgs naturalness.  Since our aim is to test how robust naturalness arguments can be let us banish further thoughts about gravity and the damage it could do to the Higgs boson and the weak scale.

Another line of thinking is to consider the prospects of many new particles at higher scales that are charged under the Standard Model gauge symmetries. For example, the existence of heavy vector-like fermions charged under the Standard Model electroweak symmetries will induce large finite quantum corrections at the two loop level (Martin 1997), and it has been argued that any fermions of this kind that exist above $10\tev$ destabilize the Higgs mass scale (Farina et al.\ 2013).  This is a powerful argument in general against the Higgs boson, since there is nothing to prevent arbitrarily heavy and arbitrary number of vectorlike fermions.  
Their masses are gauge invariant without the need of additional spontaneous symmetry breaking, and the fermions  do not contribute to gauge anomalies. Furthermore, in many string constructions there are a large number of vectorlike fermions that can arise in the spectrum. These generic aspects of vectorlike fermions are summarized recently by Ellis et al.\ (2014).  
However, assuming that the additional states have to be charged under the Standard Model for this worry to arise may seem too specialized to some.  Perhaps the underlying theory gives the Standard Model gauge groups with pure chiral fermions (i.e., left and right fermions with different charges), whose mass is then necessarily bound to the Higgs boson vacuum expectation value. There are no known vectorlike fermions in nature, and invoking their existence, giving them Standard Model gauge charges, and assuming they exist at very massive scales is maybe too much speculation to convict the Higgs boson theory and the Standard Model.

At this point we have excluded gravity and additional states charged under the Standard Model from the discussion on Higgs boson naturalness. We must ask ourselves what else could create a problem for the stability of the Higgs potential. The leading answer to this is the proliferation of additional heavy scalars in nature. 
By ``proliferation" we mean the existence of additional spin-zero scalar bosons beyond the Higgs boson that was recently discovered. All particles whose fields transform trivially (i.e., spin-zero) under the Lorentz group operations of rotations and boosts are classified as ``additional scalar bosons." The analogous categories are the spin-1/2 fermions, and the spin-1 bosons. There are at least $45$  spin-1/2 fermions in nature\footnote{The number 45 is counting colors and isospin. $Q_L$ has 6 (left-handed quarks with 2 isospins, $u_L$ and $d_L$, and each  has 3 colors), $u_R$ has 3 (right-handed up quarks with 3 colors), $d_R$ has 3 (right-handed down quark with 3 colors), $e_R$ has 1 (right-handed electron with no extra color or isospin factors), and $L$ has 2 (left-handed leptons with 2 isospins, $e_L$ and $\nu_L$). That makes a total of 15 for each of the three generation of fermions, making a total of 45 fermions.}, and even more if one counts right-handed neutrinos, and there are at least 13 spin-1 vector bosons in nature. As yet, we know of only one spin-zero scalar boson\footnote{The Higgs boson field in the Standard Model is a complex electroweak doublet with four degrees of freedom, of which three are absorbed as longitudinal components of the $Z$, $W^+$ and $W^-$ spin-1 bosons, leaving only one physically propagating scalar boson in the spectrum --- the recently discovered Higgs boson.}, the Higgs boson, and introducing more of these particles creates additional challenges that are not experienced when increasing the number of spin-1/2 and spin-1 representations. 
For example, if we introduce into the spectrum a scalar $\Phi$ with mass $M_\Phi$ one finds that there is no symmetry that forbids a renormalizable coupling between $H$ and $\Phi$ in the form of $H^\dagger H\Phi^\dagger \Phi$. 
Interactions that are not forbidden by a symmetry generically occur in quantum field theory, since a theory does not suffer from self-consistency and completeness questions ``as long as every term allowed by symmetries is included" (Weinberg 2009). 
Therefore we expect this mixing to be present. However, its presence introduces a dangerous correction to the Standard Model Higgs mass, $\Delta m_H^2\propto M^2_\Phi$. If $M^2_\Phi\gg m_H^2$ the weak scale is destabilized and wants to raise itself to the higher mass scale of $M_\Phi$.

It is this prospect of additional heavy scalars that is 
particularly 
troublesome for naturalness of the Higgs boson mass and the weak scale. In other words, it is not the intrinsic unnaturalness of the Higgs boson of the Standard Model that is necessarily so troubling, but the immediate prospects of destabilization when its kind is proliferated in nature.  There may be an intrinsic naturalness issue with the Standard Model, but that is more controversial as explained above. However, the presence of more scalars at hierarchically larger scales in nature leads to a clear instability problem. In the next section we will discuss in more detail this proliferation instability problem of the Higgs boson. We then argue that this is a real concern for the Standard Model, and that testing a theory against proliferation is not an idle speculation but is confronting a generic possibility. Finally, we show how some theories protect against proliferation instability, and in specific we show that supersymmetry is a prime example of one that solves the problem elegantly. We then end with some brief concluding comments and summary.

\section{Proliferation instability problem}

For most of its history the practitioners of particle physics have studied only particles that make up our bodies. The electrons, neutrons, protons were our first quarries. We then learned about the neutrino which is a product of nuclear beta decay, and which is a weak-interaction isospin partner with the electron. Then we learned about quarks, which are constituents of the nuclei.  We have known about and studied light and photons for much longer, which are carriers of force between our particles. We learned about the $W$ and $Z$ bosons, and also gluons, which are again force carriers between the particles that make up our bodies. It was an astonishing step when we found the muon, which opened the door to finding many new particles, which however turned out to be merely copies of particles that make up our bodies, except heavier. The discovery of the Higgs boson can be interpreted prosaically as another particle that deals with the stuff of us: it happens to give mass to the particles of our body. This is a good discovery.  Yet again, it is all about us.

Cosmologists and astrophysicists have long dealt with the notion that there is much more to the universe than what makes up our bodies. After all there is strong evidence that the stuff of us makes up only about 4\% of the energy density of the universe. Dark matter makes up another $\sim 21\%$ and dark energy $\sim 75\%$.  Many particle physicists have embraced the inevitability of dark matter (Hooper 2008), but most studies concentrate on particles that are close cousins of our ordinary particles. For example, one of the most popular dark matter candidates is a spin $1/2$ superpartner of a Standard Model gauge boson with perhaps an admixture of the spin $1/2$ superpartner of the Higgs boson (Jungman \& Kamionkowski 1996). 

On one hand it is admirable to press for maximal economy in the laws of nature. Why add more degrees of freedom to theories when a close cousin particle will do? On the other hand, it is inconceivable that nature's sole purpose is to put every particle that possibly exists at our fingertips. This implicit view that all of particle physics must be closely related to the particles that make up our human bodies is too narrow of vision. We are unlikely to be that special in the universe.

If we do not believe that we are uniquely special, then we have to ask how stable is our theory if aspects of it are 
multiplied. As we discussed earlier one of the most basic categorizations of the Standard Model is the transformation properties of the particles under the Lorentz group of rotations and relativistic boosts. We believe that nature should be invariant under these actions, and therefore particles must transform as well-defined representations of this symmetry. Indeed, we classify the electron as a spin 1/2 fermion, the photon as a spin 1 boson, and the Higgs particle as a spin 0 boson. These are all representations of the Lorentz group symmetry.

If we increase the number of fermions and vector bosons under the assumption that gravity is of no concern and that these extra states are not charged under the Standard Model gauge symmetries, there is no dramatic lifting of the weak scale hierarchy of masses. The theory is stable to such additions. There is a subtle question regarding the abelian hypercharge gauge factor, and whether abelian gauge group proliferation could destabilize the theory in a different way under some conditions, but that is to be addressed elsewhere and does not affect our argument here.

On the other hand, proliferation of condensing scalar bosons would dramatically destabilize the weak scale and the Standard Model theory. The nature of the problem is illustrated if we assume a number of other scalars similar to the Higgs boson in its broadest sense, without it being charged under Standard Model gauge symmetries. This problem we will demonstrate more concretely in the next section.

\section{Destabilization from proliferating scalars}
\label{sec:multiple}

The question we wish to investigate now is what goes wrong with the Standard Model theory when we assume that there are extra condensing scalars populating the energy landscape up to a very high scale. These extra scalars constitute a proliferation of the Higgs boson under the general category of condensing spin-zero particles. 

A proliferation of Higgs bosons charged under the Standard Model has been recognized as a serious source of concern for quite some time. Veltman (1997) has said that, 
\begin{quote}
{\it The introduction of higher Higgs multiplets, or of more than one doublet has the obvious disadvantage that in general no zero mass vector boson survives. In other words, the observed zero photon mass is then an `accident.' For this reason alone these schemes are very unattractive.}
\end{quote}
This observation speaks to the destabilization of the theory, which requires a preserved exact abelian gauge symmetry associated with electricity and magnetism. The concern of Veltman is largely when there is a proliferation of additional condensing scalar fields that are charged under the Standard Model gauge group. This may be bad enough, but let us focus less on destabilization of the theory from scalars charged under the Standard Model, and more on destabilization of the weak scale by the introduction of additional Higgs boson fields not charged under the Standard Model.  

Let us simplify the matter to begin with and assume just one additional scalar $\Phi$ that has no charges under the Standard Model gauge symmetries. Since $|\Phi|^2$ is gauge invariant and Lorentz invariant there is no prohibition to coupling it with the Standard Model Higgs boson $H$ at the renormalizable level. The resulting scalar potential is
\beq
V=-\mu_H^2|H|^2-\mu_\Phi^2|\Phi|^2+\eta |H|^2|\Phi|^2+\lambda_H |H|^4+\lambda_\Phi |\Phi|^4.
\label{eq:mixingmass}
\eeq
Assuming $\langle H\rangle=v$ and $\langle \Phi\rangle=\xi$, the minimization conditions for this potential are
\bea
-\mu_H^2+\frac{\eta}{2}\xi^2+\lambda_H v^2 &=& 0\nonumber \\
-\mu_\Phi^2+\frac{\eta}{2}v^2+\lambda_\Phi \xi^2 &=& 0
\label{eq:minimise} 
\eea 
These two equations must be satisfied to be at the stable minimum of the potential.

If we assume all dimensionless couplings are ${\cal O}(1)$ and $\mu^2_\Phi\sim \xi^2\gg v^2$, we have a serious problem with eq.~\ref{eq:minimise}. There is no reason to discount the prospect of even many condensing scalars with vacuum expectation values as high as the Planck scale, $10^{18}\gev$, which is sixteen orders of magnitude higher than the weak scale $m_{\rm weak}$, but even just this one extra field is destabilizing. Somehow the large $\mu_H^2-\eta \xi^2/2$ first two terms in the first minimization equation above must cancel each other to a large fine-tuned degree in order to match in magnitude the much smaller $\lambda v^2$ term so that the minimization condition is satisfied.  There are only two solutions to this problem. One, we accept a serious fine-tuning of the parameters such that this cancelation occurs.  Or, assume that for some reason the mixing $\eta$ between the Higgs and any other condensing scalar is small so that every term of that first equation is of the same order ${\cal O}(m_H^2)$. The mixing has to be at least as small as $\eta\sim v^2/\xi^2\ll 1$.

There are strong arguments against both solutions to this proliferation problem. And as alluded to above, the problem gets much worse as the number of condensing scalars increases.  The first solution assumes an accidental fine-tuned cancellation among terms that is hard to imagine in even just one equation. However, if we had $n$ scalars then there would be $n$ such minimization equations, all requiring similarly spectacular fine-tuned cancellations.  The small mixing solution is less than desirable also, because if there are $n$ such scalars then we have to assume that there is at least the same small mixing for every one of them. This is no longer accidental but systematic, and so must involve a principle, 
such as a symmetry or some other restriction to the theory that enforces the small mixing. 
This principle is unknown from the point of view of the Standard Model and thus is not satisfactory unless ``new physics" is invoked. 

We should remark that even non-condensing scalars when coupled to the Higgs boson, as in eq.~\ref{eq:mixingmass}, will contribute through one-loop finite quantum effects to the mass of the Higgs boson as illustrated, for example, in Martin (1997). If they coupling to the Higgs boson at ${\cal O}(1)$ strength and have mass greater than a few TeV, the Higgs mass scale is destabilized in that case as well. 

\section{Genericness of proliferation}

Another response to the proliferation problem is to assume that there simply is no proliferation of Higgs bosons in nature, and so no proliferation instability problem arises. 
One difficulty with this position, as discussed earlier, is that we would be required to believe that the Higgs boson is very special 
in that 
unlike any other representations in the Standard Model --- the spin 1/2 fermions and spin 1 vector bosons --- there is just one propagating scalar state, the Higgs boson that gives mass to the particles interacting with us, and no others. 
This position fails a modern-day Copernican test of making sure our theories do not require us to believe we are particularly special. 

Beyond these generic expectations we can be more concrete, 
and perhaps more compelling, as to the generic need for additional scalar bosons.
It is almost universally  the case that more 
complete theories that try to incorporate dark matter, inflation (Baumann 2009), flavor (Babu 2009), 
or which strive to be compatible with a theory of quantum gravity, such as string theory, generically predict that there should be many more particles and much more dynamics than just what is described by the Standard Model.  Regarding this last category, one should expect dozens, or perhaps even thousands of more Higgs bosons of exotic sectors that condense and break symmetries (Dijkstra et al.\ 2005). Since the Higgs boson is a spin-zero particle, we can immediately write a super-renormalizable (i.e., operator of dimensionality less than four) mass operator that is invariant under all gauge symmetries and spacetime symmetry, $H_i^\dagger H_i$. This in turn allows us to write down a dangerous marginal operator (i.e., operator of dimensionality equal to four) that mixes the Higgs bosons $H_i^\dagger H_iH^\dagger H$ even if $H_i$ has completely different gauge charges from the Standard Model Higgs boson $H$.  And as we saw above, this rapidly destabilizes the 
weak scale 
if the exotic scalar is heavy or if its vacuum expectation value $\langle H_i\rangle$ is parametrically larger than the weak scale, which we should think is generically possible in the presence of a collection of Higgs states at all accessible scales from the weak scale to the Planck scale.

\section{Solutions to the proliferation instability problem}

The reader may object that the discussion in sec.~\ref{sec:multiple}  supposed  values of approximately unity for the mixing parameters $\eta_k$ that mix the exotic Higgses $H_k$ with the Standard Model Higgs via $\eta_k |H_k|^2 |H|^2$, whereas in reality there may be a good reason for why the $\eta_k$ values are always suppressed by factors of at least $\sim v^2/\xi^2$. However, as we commented on above, finding the theory that enforces this requires the discovery of a new 
symmetry principle 
that goes beyond anything the Standard Model structure contemplates and reveals.  This new principle would be embedded in a new theory framework that would no longer be identified as the Standard Model.

Nevertheless, there are several solutions to the proliferation instability problem that we have been describing above. Operationally, any theory that purports to provide a solution has the burden of enforcing stability in the presence of a large number of massive or condensing scalars.  The prospective solutions include banishing  scalars (Susskind 1979) as in technicolor and composite Higgs theories, banishing high-scale hierarchies as in large extra dimensions (Arkani-Hamed et al.\ 1998) or warped extra dimensions (Randall \& Sundrum 1999), or invoking supersymmetry (Haber \& Kane 1985, Martin 1997). Not surprisingly, given our discussion above pointing out the important connection between the proliferation instability problem and naturalness, this triumvirate of general approaches that solve naturalness in its broadest formulations also can potentially solve the proliferation instability problem.  

The first solution, to banish the entire category of scalars from the theory, clearly would take care of any problem systemic to scalars. However, there are well-known challenges to matching data with this approach (Pomarol 2012), not to mention that the recent discovery of a weakly interacting Higgs boson consistent with being elementary puts strain on this idea.  

The second  solution is banishing the existence of high scales through 
extra dimensions.  The idea is to reinterpret the single number of the very large mass Planck scale of gravity as the ratio of two numbers involving the weak scale and a very large extra dimensional volume or warp factor, in the case of warped extra dimensions (Csaki 2004). This approach may not work well to solve the proliferation problem. If we indeed have dozens or more condensing scalars in nature --  let's call the number $N_H$ -- at scales not too far away from the Higgs mass scale, there is still the potential of destabilizing the  Higgs mass away from the weak scale. 
For the Higgs mass to be stable the sum of contributions to the Higgs mass-squared operator would have to be of order the Higgs boson mass, $m_H^2\sim N_H\xi^2$, where $\xi$ is the typical vacuum expectation value of the exotic condensing scalar. Thus, lowering the high-scale nearer to the weak scale through large or warped extra dimensions would soften the destabilization problem some, but may not eradicate it.

The third solution, supersymmetry, is next to consider. It is a remarkable feat of supersymmetry that the Higgs sector is generically completely stable to a large number of extra condensing Higgs bosons, in stark contrast to non-supersymmetric field theories. The key is a 
required property of supersymmetry invariance that forces interactions to be analytic in their fields. We describe in some detail in the next section how the supersymmetric solution works.

\section{The supersymmetric solution}

Within supersymmetry one can add many additional condensing scalars at any scale and no destabilizing would occur, provided two conditions are met. First, we have to assume that there is a solution to the $\mu$-problem of supersymmetry. Since supersymmetry invariance requires two Higgs doublets, $H_u$ and $H_d$, there is a Lorentz invariant and gauge invariant operator that connects the two into a single bilinear superfield interaction $H_u\cdot H_d$. 
The coefficient of this operator is $\mu$ and the concern is that there is no obvious reason why $\mu$ is of order $\mweak$ and not $M_{Pl}$. However, a small value of $\mu$ is ``technically natural",  unlike a small Higgs boson mass of the Standard Model. In other words, if it is assumed to be small there is no quantum destabilization of its value. And also, there are many elegant mechanisms by which the $\mu$ term naturally could inherit a value near $\mweak$ (Martin 1997).

The second condition is that there should not be any pure singlets in nature under all possible symmetries (Bagger \& Poppitz 1993). This condition is fine, because under almost all  general definitions of the properties of the Higgs boson a pure singlet scalar would not qualify. Also, there are no known examples in nature of a pure singlet, much less a pure singlet scalar, 
and although it would not be very convincing to hang the full argument on this fact,
it lends mild support to the supposition that this very narrow pure singlet category of nature is not allowed. 
We note that the right-handed neutrino might be a pure gauge singlet, if it exists, but it transforms nontrivially under the Lorentz group as a spinor, and in any event most theory approaches to unification of particles and forces give it charge under some other symmetry (i.e., the right-handed neutrino transforms under a non-trivial representation of the symmetry). For example, it may be part of a 16 dimensional representation of the grand unification group  $SO(10)$, and thus carry $SO(10)$ charge (Ross 2003). 
It is is therefore reasonable to suggest that nature has no pure singlets under all possible symmetries.

We therefore only consider the case where all the extra condensing scalars are charged under some symmetry or another. This restriction still lets through an instability problem because $\Phi^\dagger \Phi$ is gauge invariant anyway and destabilizes the Higgs sector when it couples to the Standard Model Higgs field operator $H^\dagger H$, as discussed earlier. Supersymmetry, on the other hand, has analytic requirements for its superpotential interactions, and therefore there is no place for $\Phi^\dagger \Phi$ to couple to the Higgs boson. This is the key to its solution. 

Let us explain further how a supersymmetric field theory absorbs  many extra scalars $\Phi_i$ charged under a variety of different symmetries, and how it preserves stability of the Standard Model Higgs boson\footnote{Supersymmetry invariance requires two Higgs boson fields, $H_u$ and $H_d$ to give mass to the up and down type fermions separately. It is a feature of supersymmetric theories that when the scale of supersymmetry breaking (i.e., the scale of exotic super partner masses) is above $\mweak$ the lightest Higgs boson is an admixture of $H_u$ and $H_u$ scalar parts and its properties are remarkably close to the Standard Model Higgs boson (Gunion \& Haber 2003).}  and the electroweak theory. Supersymmetry is a symmetry that transforms bosons into fermions and vice versa (Wess \& Bagger 1992). Supersymmetry invariance of a lagrangian among all the component fermion and boson fields of nature requires the introduction of additional boson and fermion superpartners (Haber \& Kane 1985). The addition of superpartners, effectively doubling the number of particles expected in nature, is analogous to the addition of anti-particles in standard quantum field theory, which also doubled the spectrum when introduced (Murayama 2000). The non-gauge interactions of supersymmetric theories are given in short-hand notation by a superpotential, which is an efficient and compact way of  writing down supersymmetry invariant interactions that  are allowed in the lagrangian. These interactions are computed by following a supersymmetry rulebook applied to the super potential (Wess \& Bagger 1992).

One of the results of these supersymmetry rules is that two scalars cannot couple to each other via $H^\dagger_1H_1H^\dagger_2H_2$ unless they share gauge quantum numbers with each other. However, under our considerations, this is not allowed\footnote{Even if we assumed overlapping gauge quantum numbers of the Higgs boson with some other exotic scalars,  the resulting impact on low scale theory would not destabilize the Higgs mass  even if these scalars condensed due to the existence of  stable $D$-flat directions of the potential.}. Therefore, even though $H^\dagger_1H_1H^\dagger_2H_2$ is a perfectly allowed gauge invariant operator from the standpoint of relativistic gauge theories, the analytic structure of supersymmetry will not allow it. One technicality is that the K\"ahler potential interactions can allow $H^\dagger_1H_1H^\dagger_2H_2$, but this necessarily comes with a very large suppression factor\footnote{The K\"abler interaction that would allow this is $\int d^4\theta X^\dagger X H^\dagger_1H_1H^\dagger_2H_2/M_{Pl}^4$ where $\theta$ is the Grassman variable of superspace and $X$ is the supersymmetric breaking spurion whose $F$ term is $\mweak M_{Pl}\, \theta^2$.} of $m_{\rm weak}^2/M_{Pl}^2$, and is not dangerous. 

We have found that despite the possible existence  of many heavy or condensing scalar fields, the weak scale is not destabilized if nature manifests supersymmetry not too far from $\mweak$. Supersymmetry is perhaps the most attractive theory that solves the Higgs Boson proliferation instability problem. However, it remains to be seen from experiment if nature has chosen the supersymmetry route. Confirmation would require finding spin zero super partners of the Standard Model fermions, or spin 1/2 super partners of the Higgs bosons or gauge bosons (Craig 2013). 

\section{Conclusions}

The question of whether the newly discovered Higgs boson of the Standard Model suffers from a destabilizing naturalness problem remains an important concern in particle physics. Trust in the naturalness criteria awaits the discoveries or lack of discoveries of the future higher energy proton-proton collisions  at the Large Hadron Collider to begin in earnest in 2015. In the meantime, theory support for the naturalness criteria, described as a quantum mechanical quadratic sensitivity to cutoff scales, has been questioned. 

In this article I have presented the case for a more restricted formulation of the disquieting features of the Higgs boson, avoiding considerations of gravity and avoiding the assumption of additional particles charged under the Standard Model. These are two areas of attack that naturalness has suffered recently, and it is a worthwhile study to exclude these from consideration. As a result, I have emphasized that the Higgs boson suffers from a proliferation instability problem, which I have argued should be generically addressed in all physics theories. There is reason to believe that the Higgs boson is not the only spin-zero scalar boson in nature. As soon as more are admitted a serious destabilization problem develops, as we described in the text. Theories that attempt to get by with no  scalars may solve this problem tautologically, but the discovery of the Higgs boson scalar puts this approach at risk. One would have to assume that the Higgs boson is a composite state with dynamics nearby -- both of which are disfavored by the data. Nevertheless it remains  a viable solution. Theories of extra dimensions and low-scale gravity, which banish the existence of higher destabilizing scales that the Higgs boson could couple to via other condensing scalars, are unlikely to solve the proliferation problem on its own, although they can greatly soften the problems.

Supersymmetry on the other hand appears to solve the proliferation instability problem without invoking any other features. The only requirements are that there are no pure singlet scalar states in nature, and that the technically natural $\mu$ term that connects $\mu H_u\cdot H_d$ together is near the weak scale. This is a restriction on two  narrow and technical criteria compared to the admittance of a very large number of possible exotic Higgs states charged under many different exotic symmetries. This is the proliferation concern that is most important to address, which supersymmetry passes comfortably due to its structure.

We have argued that the scalar proliferation instability problem is perhaps the least controversial, most generic and most important subcategory of the general naturalness problem. It is not unexpected that solutions to the general naturalness problem can solve or help solve the proliferation problem. Discussions of the general naturalness problem, however, have been plagued often with quasi-mystical arguments involving quantum gravity and arguments involving cancellations of bare mass terms with regulator-dependent cutoff scales, which we have no access to and are mere non-physical intermediate book-keeping devices for calculation. Considering the requirements for stability in the presence of a large number of heavy or condensing scalars in nature -- a generic and motivated consideration -- is a more concrete problem to address, and we have shown that the  requirement of proliferation stability puts significant restrictions on theory, and implies that there is more to discover beyond the Standard Model to complete our understanding of the weak scale.








\bigskip\bigskip

\noindent
{\bf References}
\bigskip

\parindent=0in

\bibentry{Aad, G.\ et al.\ (ATLAS Collaboration). (2012).\ Observation of a New Particle in the Search for the Standard Model Higgs Boson with the ATLAS Detector at the LHC. {\it Phys.\ Lett.}\ B716, 1-26 [arXiv:1207.7214].}

\bibentry{Arkani-Hamed, N., Dimopoulos, S., Dvali, G.\ (1998). The hierarchy problem and new dimensions at a millemeter. {\it Phys. Lett.} B429, 263-272.}

\bibentry{Babu, K.S. (2009).\ Lectures on Flavor Physics. arXiv:0910.2948 [hep-ph].}
  
\bibentry{Bagger, J.\ \& Poppitz, E.\ (1993). Destabilizing divergences in supergravity coupled supersymmetric theories. Phys.\ Rev.\ Lett.\ 71, 2380.}

\bibentry{Bardeen, W.A. (1995). On naturalness in the standard model. {\it Proc. Ontake Summer Inst.\ on Particle Physics}, 27 Aug - 2 Sep 1995, Ontake Mountain, Japan, Fermilab-Conf--95-391-T.}

\bibentry{Baumann, D.\ (2009). Lectures on Inflation. {\it Proceedings of the Theoretical Advanced Study Institute}, Boulder. arXiv:0907.5424.}

\bibentry{Chatrchyan, S.\ et al.\ (CMS Collaboration). (2012).\ Observation of a new boson at a mass of 125 GeV with the CMS experiment at the LHC. {\it Phys.\ Lett.}\ B716, 30-61 [arXiv:1207.7235].}

\bibentry{Craig, N.\ (2013). The state of supersymmetry after run I of the LHC. {\it Proceedings of the 11th Conference on Flavor Physics and CP Violation (FPCP 2013)}. 20-24 May 2013, Rio de Janeiro, Brazil.}

\bibentry{Csaki, C.\ 2004. Extra dimensions and branes. In {\it From fields to strings, vol.\ 2}, ed.\ M.\ Shifman, World Scientific.}

\bibentry{Dijkstra et al.\ 2005.\ Supersymmetric Standard Model spectra from RCFT orientifolds. Nucl.\ Phys.\ B710, 3.}

\bibentry{Drees, M., Godbole, R.M, Roy, P.\ (2004). {\it Theory and Phenomenology of Sparticles.} World Scientific.}

\bibentry{Ellis, S.A.R, Godbole, R.M., Gopalakrishna, S., Wells, J.D.\ (2014). Survey of vector-like fermion extensions of the Standard Model and their phenomenological implications. To be published in {\it JHEP}. arXiv:1404.4398}

\bibentry{Farina, M., Pappadopulo, D., Strumia, A.\  (2013). Modified naturalness principle and its experimental tests. {\it JHEP} 1308, 022, arXiv:1303.7244.}

\bibentry{Giudice, G. (2008).\ Naturally Speaking: The Naturalness Criterion and Physics at the LHC. In G. Kane, A. Pierce (eds.), {\it Perspectives on LHC Physics}, 2008 [arXiv:0801.2562].}
  
\bibentry{Giudice, G.\ \& Romanino, A.\ (2004).\ Split supersymmetry. {\it Nucl.\ Phys.}\ B699, 65 (2004) [hep-ph/0406088].}

\bibentry{de Gouv\^ea, A., Hern\'andez, D., Tait, T.M.P.\ (2014). Criteria for Natural Hierarchies. arXiv:1402.2658.}

\bibentry{Gunion, J.F.\ \& Haber, H.E.\ (2002). The CP conserving two Higgs doublet model: the approach to the decoupling limit. {\it Phys.\ Rev.} D67, 075019.}

\bibentry{Haber, H.E. \& Kane, G.L.\ (1985). The search for supersymmetry: probing physics beyond the Standard Model. {\it Phys.\ Rep.} 117, 75.}

\bibentry{`t Hooft, G. (1980).\ In {\it Recent Developments in Gauge Theory.} New York: Plenum Press.}

\bibentry{Hooper, D.\ (2009). TASI 2008 Lectures on Dark Matter. {\it Proceedings of the Theoretical Advanced Study Institute}, Boulder. arXiv:0901.4090.}

\bibentry{Jungman, G., Kamionkowski, M.\ (1996). Supersymmetric dark matter. {\it Phys.\ Rep.}\ D51, 3121.}

\bibentry{Martin, S.P. (1997). Supersymmetry Primer. Published in {\it Perspectives on Supersymmetry}, World Scientific, 1998.}

\bibentry{Murayama, H. (2000).\ ``Positron Analogue" in Supersymmetry Phenomenology. {\it Proceedings of 1999 ICTP Summer School.} arXiv:hep-ph/0002232.}

\bibentry{Pomarol, A.\ (2012). Beyond the Standard Model. {\it Lectures at the 2010 European School of High-Energy Physics}, 20 June -- 3 July 2010, Raseborg, Finland. arXiv:1202.1391.}

\bibentry{Randall, L. \& Sundrum, R.\ (1999). A large mass hierarchy from a small extra dimension. {\it Phys. Rev. Lett.}\ 83, 3370.}

\bibentry{Rugh, S.E.\ \& Zinkernagel, H.\ (2002).\ The Quantum Vacuum and the Cosmological Constant Problem.
{\it Studies in the History and Philosophy of Modern Physics} {\bf 33}, 663-705.}

\bibentry{Ross, G.G.\ (2003). {\it Grand Unified Theories.} New York: Westview Press.}
  
\bibentry{Soni, N.\ (2013). Searches for Physics Beyond the Standard Model at the LHC with the ATLAS Detector. {\it Proceedings of the Sixth High-Energy Physics International Conference}, HEP-MAD 13. Antananarivo, Madagascar (4-10 September 2013).}

\bibentry{Susskind, L.\ (1979). Dynamics of spontaneous symmetry breaking in the Weinberg-Salam theory. {\it Phys.\ Rev.\ D}20, 2619.}

\bibentry{Veltman, M.J.G. (1997). {\it Reflections on the Higgs System.} CERN Yellow Report, volume 97, issue 05.}

\bibentry{Weinberg, S.\ (2009). Effective Field Theory, Past and Future. {\it Proc.\ of  6th Intl.\ Workshop on Chiral Dynamics}. Bern, Switzerland, 6 July 2009. [arXiv:0908.1964]}
  
\bibentry{Wess, J. \& Bagger, J.\ (1992). {\it Supersymmetry and Supergravity.} Princeton University Press.}

\end{document}